# The origin of the spontaneous electric polarization

## Shpend Mushkolaj

e-mail: sh.mushkolaj@hispeed.ch

## Abstract

The phenomenon of the spontaneous electric polarization (ferro- and antiferroelectricity) is one of the fundamental problems of the solid-state physics. Although, there has been lot of experimental and theoretical progress, still needs to be made to understand the origine of the spontaneous electric polarization. A new microscopic theory that is based on the perfectly elastic electron-atom and atom-atom collisions, crystal structure and the chemical composition of the compounds is developed to calculate the critical temperatures for materials with spontaneous electric polarization. The reliability of this model was examined by comparing the experimental results with calculated values.

## 1. Introduction

The spontaneous parallel and anti-parallel orientations of the electric dipole moments are called as ferroelectricity or antiferroelectricity, respectively. The first parallel orientation of the dipole moments was discovered in 1920[1] in the crystal of the Rochelle Salt. Since then many different compounds that show a spontaneous electric polarization have been found. Utilization of these materials in optics, acoustics, capacitor engineering and computer technology is essential. Although, there is lot of theoretical progress in the understanding of the nature of spontaneous electric polarization, still is missing a microscopic theory that is based only on the crystal structure and chemical composition of the compounds. Such a theory should be able to predict what kind of elements are needed to combine to create ferro- and antiferroelectric materials with a desired critical temperature. It should also be able to calculate the critical temperatures for both types of the spontaneous polarized materials, namely the order-disorder and displacive type. At this point it is worth to remember that in the order-disorder type the spontaneous polarization is caused by the displacements of ions that do not belong to the ordered sublattice, while in the displacive type it is originated by displacements of ions that belong to the ordered sublattice

Before presenting the new microscopic theory it is important to show the actual most accurate formula for calculating the critical temperature $T_c$, that is derived from the mean field theory:

$$T_c = C_{C-W} \frac{J_0 V_{uc}}{4\pi p_z^2} \qquad (1)$$

where $C_{C-W}$, $J_0$, $V_{uc}$ and $p_z$ represent the Curie-Weiss constant, the interaction of a particle with all other particles being located within a large interaction radius (*i.e.* $r_0 >>$ lattice parameters), volume of the unit cell and the dipole moment, respectively. It should be pointed out that the Eq. (1) is valid only for ferroelectrics of the order-disorder type.

For the materials that are classified as belonging to the displacive type the formula for $T_c$ is different form the Eq. (1)[2] and it is inaccurate to predict the critical temperatures.

In Eq. (1) the quantity $J_0 V_{uc}/p_z^2 = \beta$ is known as the Lorentz factor, and for the cubic crystals with dipole located in the center of the unit cell, this factor is, $\beta = 4\pi/3$. After inserting the Lorentz factor the Eq. (1) transforms into:

$$T_c = \frac{C_{C-W}}{3}. \qquad (2)$$

In the table I it is shown that for materials belonging to the order-disorder type, Eq. (2) is fulfilled only in order of magnitude, while for materials of displacive type it is not fulfilled:

Table I: These data have been taken from the Ref. [2].

| Compounds | type | $T_c^{exp.}$ (K) | $C_{C-W}$ (K) |
|---|---|---|---|
| TGS | order-disorder | 322 | 3200 |
| NaNO$_2$ | order-disorder | 437 | 5000 |
| KH2PO$_4$ | order-disorder | 123 | 3600 |
| BaTiO3 | displacive | 400 | 170000 |

For hydrogen-containing ferroelectrics there are many formulas for calculating the critical temperatures, but the agreement with the experimental data is very poor. It is worth to point out that in general the formula for calculating the $T_c$ is derived from the minimization of the free energy with the respect of the order parameters.

In this paper is presented a novel formula for calculating the $T_c$, where $T_c$ depends only on the crystal structure and on the chemical compositions of the compounds. The formula for the $T_c$ will be derived from the solution of the time-dependent Schrödinger Equation, where collisions between electrons and the atoms that cause the spontaneous polarization are perfectly elastic, *i.e.* the kinetic energy is conserved. To prove the reliability of the novel formula for $T_c$ a comparison of the calculated values with experimental results for different materials will be presented.

## 2. The derivation of the formula for critical temperatures

The main focus in this section will be on the derivation of the general formula for critical temperatures. The $T_c$ should depend only on the crystal structure (*i.e.* the lattice parameters and atom-atom distances) and the chemical composition (*i.e.* the atomic masses) of the compounds. The validity of the formula should be universal in the sense that it should be able to calculate the $T_c$-s for order-disorder and displacive type materials.

Firs it will be supposed that at and below the critical point there exist some particular collisions between electrons and atoms where no kinetic energy is dissipated into the heat, i.e. that these collisions are perfectly elastic. In some multicomponent compounds it could occur that in addition to the perfectly elastic collisions (*pec*) between electrons and atoms there exist also *pec* between atoms with different masses. The kinetic energies and momentum for electrons and atoms are given by $E_{ke} = p_e^2/2m_e$, $E_{ka} = p_a^2/2M$, $p_e = m_e V_e$ and $p_a = m_e V_e$, respectively. To have a long-range effect it will be also supposed that before, during and after *pec* the electron and atom kinetic energies remain constant, i.e. $E_{ke(before\ collision)} = E_{ke(after\ collision)} = E_{ka(before\ collision)} = E_{ka(after\ collision)} = E_{kae(during\ collision)}$. It is understandable that at any time also the total kinetic energy, $(E_{ke} + E_{ka})$ remains constant. Because collisions between electrons and atoms are perfectly elastic the total momenta before and after collision time must remain also constant, i.e., $(|p_e| + |p_a|)_{(before\ collision)} = (|p_e| + |p_a|)_{(after\ collision)}$, where $|p_e|$ and $|p_a|$ represent the absolut values of the electron and atom momenta before and after *pec*, respectively.

From the kinetic energy conservation during collision time, one can easily create a free

„particle" with a kinetic energy of $E_{kae}=p_ap_e/2(m_eM)^{1/2}$ that is equal to $E_{ke}$ and $E_{ka}$, i.e., $p_ap_e/2(m_eM)^{1/2} = p_e^2/2m_e = p_a^2/2M$.

In this case the corresponding kinetic term of the Hamiltonian operator for the free „particle" with the mass of $(m_eM)^{1/2}$ may be expressed as:

$$\hat{H} = \frac{-\hbar^2}{2\sqrt{Mm_e}} \frac{\partial^2}{\partial x^2}. \tag{3}$$

The time-dependent Schrödinger wave equation is given by:

$$\frac{-\hbar^2}{2\sqrt{Mm_e}} \frac{\partial^2}{\partial x^2} \Psi(x,t) = i\hbar \frac{\partial}{\partial t} \Psi(x,t). \tag{4}$$

After inserting the plane wave function of the form:

$$\Psi(x,t) = e^{i(\mp kx - \frac{4\Delta E t}{h})}, \tag{4a}$$

into the time-dependent Schrödinger equation (4) one get for $\Delta E$:

$$\Delta E = \frac{h\hbar}{8\sqrt{M m_e}} k^2. \tag{5}$$

for $k = \pi/a$ and $\Delta E = k_B T_c$ one get the „magic" formula for $T_c$:

$$T_c = \frac{\pi h^2}{4 k_B} \frac{1}{\sqrt{M m_e} a^2}, \tag{6}$$

where *a* can be one of the lattice parameters, or other distances such as the diagonal length or the minimal distance between atoms that collide with each other with *pec*. As one can see it depends only on the lattice parameters or on the minimal atom-atom distances and on the mass of the atoms which are responsible for the spontaneous polarization.

In the multicomponent compounds where more than one type of atoms could participate on the spontaneous electric polarization, the mass of the free „particle" may be expressed in different forms. In these case in addition to the electron-atom *pec* also the atom-atom *pec* should be taken into account. Depending on that, if a dyad or a triad of atoms with masses $M_A$, $M_B$ and $M_C$ participates on the spontaneous polarization, than from the kinetic energy conservation one get the respective expressions for masses of the free „particles":

$$\sqrt{\sqrt{M_A M_B} m_e}_{(DYAD)} \quad \text{and} \quad \sqrt{\sqrt{M_C \sqrt{M_A M_B}} m_e}_{(TRIAD)}. \tag{7}$$

In these cases the Eq. (6) will be transformed into the respective forms:

$$T_c = \frac{\pi h^2}{4 k_B} \frac{1}{\sqrt{\sqrt{M_A M_B} m_e} a^2} \quad \text{and} \quad T_c = \frac{\pi h^2}{4 k_B} \frac{1}{\sqrt{\sqrt{M_C \sqrt{M_A M_B}} m_e} a^2}. \tag{8}$$

From the Eq. (4a) one can see that ($\psi(x,t)$, $\psi(-x,t)$) are the eigenstates of the time-dependent Schrödinger equation.

# 3. The application of the formula for critical temperatures

The purpose of this section is to test the reliability of the „magic" formula for $T_c$. It will be also shown the formula is universal in the sense that it is able to calculate the $T_c$-s for order-disorder and displacive type materials. In the table II the calculated values with experimental results for $T_c$-s are compared.

Table II: Comparison between calculated and experimental data for $T_c$-s.

| Chemical composition | The mass of the free „particle" | $a$ (Å) | $T_c^{calc.}$ (K) | $T_c^{exp.}$ (K) |
|---|---|---|---|---|
| $K_2SeO_3$ | $(M_K m_e)^{1/2}$ | $c=10.47$ | 93 | 93[3] |
| $KH_2PO_4$ | $[(M_K M_O)^{1/2} m_e]^{1/2}$ | $(a^2+c^2)^{1/2}= 10.18$ $a=7.444;\ c=6.945$[4] | 123 | 123[5] |
| $KD_2PO_4$ | $(M_K m_e)^{1/2}$ | $c=6.945$ | 113 | 113[5] |
| $Ba_{4.13}Na_{1.74}Nb_{10}O_{30}$ | $(M_{Na} m_e)^{1/2}$ | $c=3.9949$ | 836 | 833[6] |
| $CsH_2AsO_4$ | $[(M_{Cs} M_O)^{1/2} m_e]^{1/2}$ | $a=7.9852$ | 147 | 143[7] |
| $BaTi_2O_5$ | $[(M_{Ti} M_O)^{1/2} m_e]^{1/2}$ | $b=3.943$ | 782 | 753[8] |
| $NaNO_2$ | $(M_{Na} m_e)^{1/2}$ | $c=5.56$ | 432 | 437[9] |
| $PbTiO_3$ | $[(M_{Ti} M_O)^{1/2} m_e]^{1/2}$ | $c=4.14$ | 706 | 700[10] |
| $SrBi_3Ti_2NbO_{12}$ | $(M_{Nb} m_e)^{1/2}$ | $a=b=3.85$ | 448 | 441[11] |
| $PbBi_3Ti_2NbO_{12}$ | $(M_O m_e)^{1/2}$ | $(a^2+b^2)^{1/2}= 5.4659$ $a=b=3.865$ | 535 | 570[11] |
| $PbHfO_3$ | $[(M_{Pb} M_O)^{1/2} m_e]^{1/2}$ | $a=4.136$ | 492 | 488[12] |
| $[C_3N_2H_5]_5[Bi_2Br_{11}]$ | $[(M_{Bi} M_{Br})^{1/2} m_e]^{1/2}$ | $d_{(Bi-Br1-Bi1)}=6.021$ | 155 | 155[13] |
| $TlH_2PO_4$ | $(M_{Tl} m_e)^{1/2}$ | $a=4.525$[5] | 221 | 230[5] |
| $TlD_2PO_4$ | $(M_O m_e)^{1/2}$ | $c=6.526$[5] | 366 | 353[5] |
| $LiNbO_3$ | $(M_{Li} m_e)^{1/2}$ | $2*(R_{Li}+R_O)=4.1$ | 1446 | 1480[14] |
| $LiTiO_3$ | $(M_O m_e)^{1/2}$ | $2*(R_{Li}+R_O)=4.1$ | 951 | 950[14] |
| $RbH_2PO_4$ | $\{[(M_{Rb}(M_{Rb}M_O)^{1/2}]^{1/2} m_e\}^{1/2}$ | $a=7.608$ [3] | 147 | 147[15] |
| $RbH_2AsO_4$ | $\{[(M_{Rb}(M_{Rb}M_O)^{1/2}]^{1/2} m_e\}^{1/2}$ | $a'=R_{As}/R_P*a=8.7492$ | 111 | 110[16] |
| $BaTiO_3$ | $(M_{Ba} m_e)^{1/2}$ | $d_{min(Ba^{+2}-Ba^{+2})}=3.663$ | 407 | 403[17] |
| $BaTiO_3$ | $(M_{Ti} m_e)^{1/2}$ | $(a^2+b^2)^{1/2}= 5.678$ $a=3.99;\ b=4.04$ | 282 | 278[17] |
| $BaTiO_3$ | $[(M_{Ba} M_O)^{1/2} m_e]^{1/2}$ | $(a^2+b^2)^{1/2}= 5.678$ $a=3.99;\ b=4.04$ | 284 | 278[17] |
| $BaTiO_3$ | $(M_{Ti} m_e)^{1/2}$ | $d_{[111]}=(3)^{1/2}*4.04=6.997$ | 188 | 183[17] |
| $BaTiO_3$ | $[(M_{Ba} M_O)^{1/2} m_e]^{1/2}$ | $d_{[111]}=(3)^{1/2}*4.04=6.997$ | 189 | 183[17] |
| $[C_3N_2H_5]_5[Bi_2Cl_{11}]$ | $[(M_{Cl} M_N)^{1/2} m_e]^{1/2}$ | $c=9.045$ | 166 | 166[18] |
| $Bi_2NiMnO_6$ | $\{[(M_{Mn}(M_{Bi}M_{Mn})^{1/2}]^{1/2} m_e\}^{1/2}$ | $a=3.8775$ | 486 | 485[19] |

As one can see from the listed compounds in the above table there are all types of compounds with spontaneous electric polarization. This formula gives us also the possibility to identify the atoms that participate on the electric polarization. The identifying of the atoms that contribute to the electric polarization may be very useful for the material scientists, where a tuning of the $T_c$ through the changing of the chemical composition may be controllable.

References:

1) J. Valasek: Phys. Rev. **15** (1920) 537.
2) B. A. Strukov and A. P. Levanyuk: *Ferroelectric Phenomena in Crystals*, Springer, Madrid, Moscow, 1997.
3) E. Y. Tonkov: *High Pressure Phase Transformations,* **2** (1992) 515, Gordon and Breach Science Publishers, Philadelphia.
4) B. Dam, P. Bennema, and W. J. P. Van Enckevort, J. Cryst. Growth **74** (1986) 118.
5) S. Rios: These de Doctorat de l'Universite Paris VI (1997).
6) P.B. Jamieson, S.C. Abrahams and J.L. Bernstein: J. Chem. Phys. **50** (1969) 4352-4363 .
7) R. S. Katiyar, J. F. Ryan, and J. F. Scott: Phys. Rev. B **4** (1971) 2635 - 2638.
8) Y. Akishige and H. Shigematsu: J. Kor. Phys. Soc., **46** (2005) 24-28.
9) P. Ravindran, A. Delin, B. Johansson, O. Eriksson and J. M. Wills: Phys. Rev. B **59** (1999) 1776.
10) Young-Han Shin, Jong-Yeog Son, Byeong-Joo Lee, Ilya Grinberg and Andrew M Rappe: J. Phys.: Condens. Matter **20** (2008) 015224.
11) I. A. Trifonov, G. A. Geguzina, E. S. Gagarina, V. D. Komarov, A. V. Leyderman, E. T. Shuvaeva, A. T. Shuvaev and E. G. Fesenko: Inorganic Materials **36** (2000) 183-187.
12) G. Shirane and R. Pepinsky: Phys. Rev. **91** (1953) 812 - 815.
13) A Piecha, A Białońska and R Jakubas: J. Phys.: Condens. Matter **20** (2008) 325224.
14) I. Inbar and R. E. Cohen: Phys. Rev. B **53** (1996) 1193 - 1204.
15) Y. Le Grand, D. Rouede, C. Odin, R. Aubry and S. Mattauch: Optics Comm.*,* **200** (2001) 249-260.
16) M. A. Pimenta: Phys. Rev. B **57** (1998) 22 - 24.
17) Y. Yoshimura, T. Koganezawa, S. Morioka, H. Iwasaki, A. Kojima and K. K. Tozaki: Acta Cryst. **A61**, (2005) C324.
18) R. Jakubas, A. Piecha, A. Pietraszko and G. Bator: Phys. Rev. B **72**, (2005) 104107.
19) P. Padhan, P. LeClair, A. Gupta and G. Srinivasan: J. Phys.: Condens. Matter **20,** (2008) 355003.